\documentclass[twocolumn]{aastex6}
\usepackage{amsmath}
\usepackage{bm}
\usepackage{graphicx}
\usepackage{amssymb}
\usepackage{array}
\usepackage{color}

\begin{document}

\title{Pebble accretion at the origin of water in Europa}

\author{Thomas Ronnet, Olivier Mousis, and Pierre Vernazza}

\affil{Aix Marseille Univ, CNRS, LAM, Laboratoire d'Astrophysique de Marseille, Marseille, France  {\tt thomas.ronnet@lam.fr}}

 
\begin{abstract} 

 Despite the fact that the observed gradient in water content among the Galilean satellites is globally consistent with a formation in a circum-Jovian disk on both sides of the snowline, the mechanisms that led to a low water mass fraction in Europa ($\sim$$8\%$) are not yet understood. Here, we present new modeling results of solids transport in the circum-Jovian disk accounting for aerodynamic drag, turbulent diffusion, surface temperature evolution and sublimation of water ice. We find that the water mass fraction of pebbles (e.g., solids with sizes of 10$^{-2}$ -- 1 m) as they drift inward is globally consistent with the current water content of the Galilean system. This opens the possibility that each satellite could have formed through pebble accretion within a delimited region whose boundaries were defined by the position of the snowline. This further implies that the migration of the forming satellites was tied to the evolution of the snowline so that Europa fully accreted from partially dehydrated material in the region just inside of the snowline.

\end{abstract}

\keywords{ planets and satellites: formation  -- planets and satellites: individual (Jupiter, Galilean satellites)  -- protoplanetary disks -- methods: numerical }

\section{Introduction}

The four Galilean satellites (Io, Europa, Ganymede and Callisto) are thought to have formed during the very late stages of Jupiter's formation, at a time when Jupiter was surrounded by a circumplanetary disk (CPD) \citep[see e.g.,][]{can09,est09}. While of comparable masses, these four satellites have different densities \citep[Io: $3527.5\pm 2.9$ kg m$^{-3}$, Europa: $2989\pm 46$ kg m$^{-3}$, Ganymede: $1942.0\pm 4.8$ kg m$^{-3}$, Callisto: $1834.4\pm 3.4$ kg m$^{-3}$;][]{schub04} due to different water mass fractions (Io: $\sim$0\%, Europa: $\sim$8\%, Ganymede and Callisto: $\sim$50\%) and their density decreases (hence their water mass fraction increases) with increasing distance to Jupiter \citep{and98,sohl02,schub04}. This gradient in water mass fraction puts a strong constraint on (1) the satellites formation conditions and/or (2) their subsequent thermal evolution via tidal heating. 

Concerning case (2), it has been proposed that the density gradient among the satellites results from increased tidal heating \citep{can09,dwy13} with decreasing distance from the planet. However, Io is currently dissipating $\sim$1 ton/s of material in the Jovian magnetosphere which, integrated over 4 billion years, represents only $\sim$0.1\% of its mass. This argument alone is insufficient to fully preclude the proposed mechanism but it suggests that tidal heating is not the most likely mechanism to explain Io's or Europa's low water content \citep[see also the discussion in][]{can09}.

Concerning case (1), a first explanation that has been proposed is an increasing relative velocity among the building blocks with decreasing distance from the planet leading to substantial water loss in the case of the most energetic impacts \citep{est06} which occurred closer to Jupiter. Nonetheless, this scenario has been discarded by a detailed study showing that Io and Europa analogues exhibit an overabundance of water when they are formed via an $N$-body code simulating imperfect accretion and water loss during collisions \citep{dwy13}. A second explanation is that the observed water gradient among the satellites results from an outwardly decreasing temperature of the CPD, leading to the existence of a snowline at a given radial distance from Jupiter \citep[see e.g.,][]{lun82}. In this case, bodies that formed inward of the snowline (Io) accreted from essentially water-poor building blocks whereas bodies that formed outward of the snowline (Ganymede, Callisto) formed from a primordial mixture of water ice and silicates \citep[e.g.,][]{can02,mos03a,mos03b,MG04}. Within this scenario, the low water content of Europa is puzzling. So far, Europa's water content has been mostly attributed to its formation both outward and inward of the snowline due to either i) its migration inward of the snowline during formation (i.e. growth), ii) the progressive cooling of the disk and thus inward migration of the snowline during its formation, or iii) an interplay between the two mechanisms \citep{alib05,can09}. 
However, the evolution of the CPD has been systematically modeled using an ad-hoc parametrization of turbulent viscous disk \citep[the so-called $\alpha$-viscosity,][]{Sh73} which governs the temperature evolution and lifetime of the disk. 
While providing a good starting point for evolutionary disk models, this kind of parametrization has been highly questioned in the recent years \citep{bai13,sim13,gre15}. Hence, using a predefined $\alpha$-viscosity prescription to describe the CPD's evolution and provide hints on Europa's formation remains questionable. 
The same remark holds for planet (or satellite) migration, which has also been extensively studied within the recent years \citep[see e.g.,][]{paar10,bits14}. These studies have shown that in realistic disk conditions, migrating planets can behave significantly differently from what was previously thought, i.e. a persistent inward motion \citep[e.g.][]{tana02}, due to the existence of regions where the migration is halted and even reversed. Because the studies of satellites formation were based so far on the migration formulation of \citet{tana02} \citep[e.g.][]{can02,can06,alib05,sas10} their proposed growth/migration scenario is questionable.

Overall, it appears that Europa's composition (as well as those of the others Galilean moons) is the consequence of the way the satellite formed within the Jovian CPD rather than the result of some post-formation mechanism. Hence, investigating how the partial devolatilization of Europa's building blocks occurred within the circum-Jovian disk should provide important constraints on the processes that took place during its formation.
 
In this work, we investigate the formation conditions of the Galilean moons, and in particular those of Europa, by coupling a transport model of solids including aerodynamic drag, turbulent diffusion, surface temperature evolution and water ice sublimation with a classical  CPD prescription. Considering the fact that \citet{dwy13} demonstrated the inability of classical accretion of large (D$\sim$10--100 km) satellitesimals to reproduce the observed density gradient among the satellites, we focus here on the evolution of the so-called pebbles (D$\sim$1cm--10 m). Pebble accretion has become an attractive scenario over recent years as it is able to explain the growth of both the planets and the small bodies in our solar system \citep[see e.g.,][]{lam12,lam14a,morb15}. 

The outline of our paper is as following. The transport model of solids and the used CPD prescription are detailed in Section~\ref{meth}. The results of our simulations are presented in Section~\ref{res}. Sections ~\ref{disc} and~\ref{end} are devoted to discussion and conclusions, respectively.
  
\section{Methods}

 In this section, we provide a detailed description of our model. Similarly to \citet{can02} and \citet{sas10}, we used a simple quasi-stationary CPD model to i) derive the gas density and temperature distributions and ii) analytically determine the radial and azimuthal velocities of the gas (Section~\ref{CPD}. To model the transport of solids (Section~\ref{PDT}), we numerically solved the equation of motion of the solid particles, including the effect of gas drag, turbulent diffusion and sublimation of water ice.

\label{meth}
\subsection{Circumplanetary disk model}
\label{CPD}

The gas surface density of our CPD is derived from the gas starved disk model of \citet{can02}. In this concept, the CPD is fed through its upper layers from its inner edge up to the centrifugal radius $r_c$ by gas and gas-coupled solids inflowing from the protosolar nebula (PSN). In practice, the centrifugal radius, which corresponds to the location where the angular momentum of the incoming gas  is in balance with the gravitational potential of Jupiter, evolves with time and moves toward larger distances with respect to the growing Jupiter. 

Here, we focused on the very late stages of Jupiter's formation when the satellites start their accretion. We thus considered the centrifugal radius at a fixed distance $r_c = 26 R_\mathrm{Jup}$ for all our simulations \citep[see e.g.,][]{mos03a,can02,sas10}. The surface density is obtained by considering an equilibrium between the mass inflowing from the PSN onto the CPD and the mass accretion rate $\dot{M}_p$ onto Jupiter \citep{can02}:

\begin{equation}
\Sigma_g (r) = \frac{\dot{M}_p}{3\pi\nu(r)} \left \lbrace \begin{array}{ll}
1 - \frac{4}{5} \sqrt{\frac{R_c}{R_d}} - \frac{1}{5} \left(\frac{r}{R_c}\right)^2 & \mathrm{for} \; r\leqslant R_c\\
\\
\frac{4}{5} \sqrt{\frac{R_c}{r}} - \frac{4}{5} \sqrt{\frac{R_c}{R_d}} & \mathrm{for} \; r > R_c,\\
\end{array} \right.
\end{equation}

\noindent where $R_d$ is the outer radius of the disk, here assumed to be equal to 150$\;R_\mathrm{Jup}$ based on 3D hydrodynamic simulations \citep{tan12}. $\nu$ is the turbulent viscosity given by \citep{Sh73}: 

\begin{equation}
\nu = \alpha H^2_g \Omega_K,
\end{equation}

\noindent where $\alpha$ is the coefficient of turbulent viscosity, $\Omega_K$ the keplerian orbital frequency. $H_g = c_g / \Omega_K$ is the gas scale height derived from the isothermal gas sound speed $c_g = \sqrt{R_g T_d / \mu_g}$. $R_g$ is the ideal gas constant, $\mu_g$ the mean molecular weight of the gas ($\sim$2.4 \,g/mol), and $T_d$ the CPD's temperature at a given distance from the planet. The temperature profile is derived from the simple prescription of \citet{sas10}:

\begin{equation}
T_d \simeq 225 \left( \frac{r}{10 \, R_\mathrm{Jup}} \right)^{-3/4} \,  \left( \frac{\dot{M}_p}{10^{-7} \, M_\mathrm{Jup}\,\mathrm{yr}^{-1}} \right)^{1/4} \; \mathrm{K} . 
\end{equation}

\noindent This temperature profile is obtained from the balance between the energy provided by viscous dissipation within the CPD and the energy loss through blackbody radiation of the disk. This expression gives the temperature at the radiative surface of the disk, where energy balance is achieved. The temperature at the midplane of the disk $T_m$ is obtained by multiplying $T_d$ by a factor $(\frac{3\tau_R}{8}+\frac{1}{2\tau_P})^{1/4}$ \citep[e.g.,][]{HG05}, where $\tau_R$ and $\tau_P$ are the Rosseland and Planck mean optical depths, respectively. This would give a slightly higher temperature than $T_d$. However, given the uncertainties on the opacity, the turbulence level and the mass accretion rate of the circumjovian disk, we follow \citet{sas10} in adopting $T_m \sim T_d$. Both the surface density and gas temperature are thereby determined from the value of the accretion rate $\dot{M}_p$. Therefore, a time evolution of the CPD can be accounted for by imposing a decrease of the mass accretion rate over time. Following \citet{sas10}, this can be expressed as
\begin{equation}
  \dot{M}_p(t) =  \dot{M}_{p,0} \, e^{-\frac{t}{\tau_\mathrm{disk}}}
\end{equation}

\noindent where $\dot{M}_{p,0}$ is the initial mass accretion rate and $\tau_\mathrm{disk}$ is the lifetime of the nebula which drives its evolution.

\begin{figure*} 
\includegraphics[scale=0.45]{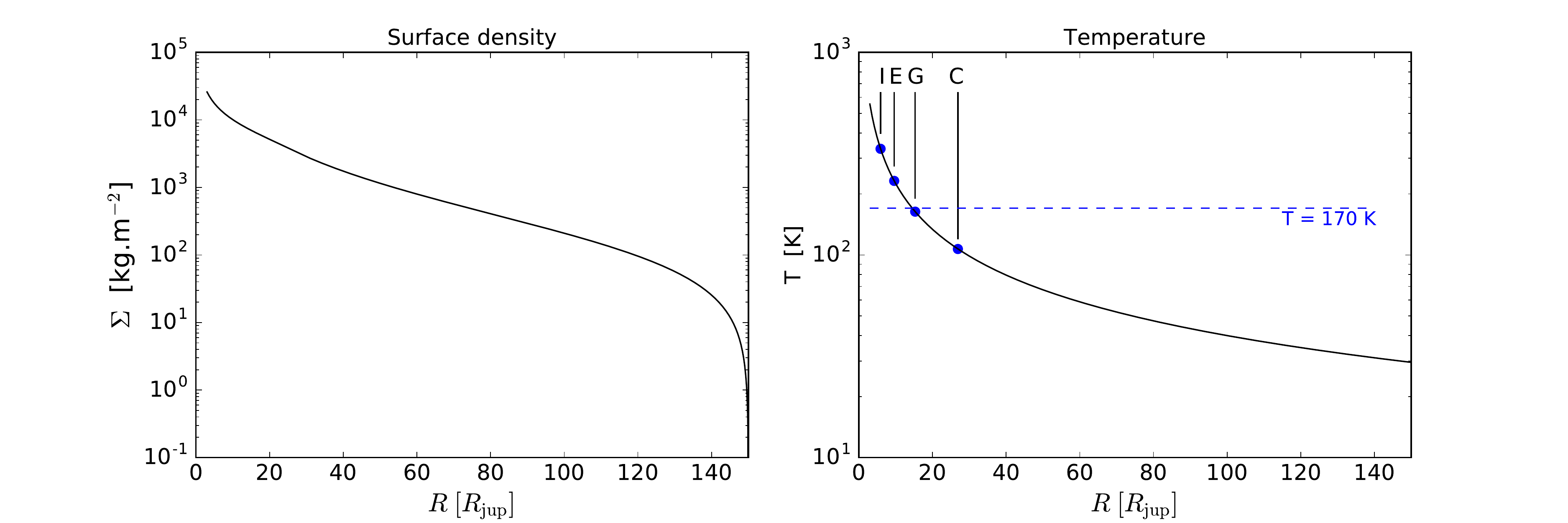}
\caption{Surface density and temperature profiles of the CPD, with the distance from Jupiter expressed in units of Jovian radii ($R_{\rm Jup}$) calculated for $\dot{M}_p = 1 \times 10^{-7} \, M_\mathrm{Jup}\, yr^{-1}$ and $\alpha = 10^{-3}$. The vertical bars designated by the letters I, E, G and C correspond to the current orbits of Io, Europe, Ganymede and Callisto, respectively.}
\label{subneb}
\end{figure*}

\noindent An example of surface density and temperature profiles of the CPD is presented in Figure~\ref{subneb} for a mass accretion rate of $10^{-7}\,M_\mathrm{Jup}\,yr^{-1}$ and a turbulent parameter $\alpha = 10^{-3}$. In that case, the temperature profiles allow the survival of water ice at Ganymede and Callisto's current location. 

Because this work aims at describing the interaction between solid particles and the gas, we have added a prescription computing the velocity of the CPD's gas. To do so, we considered that the gas is in hydrostatic equilibrium in the vertical direction and the vertical velocity of the gas is therefore zero (see \citet{tak02} for a discussion about the validity of this assumption). In the radial direction however, the generally outward pressure gradient force causes the gas to rotate at a slightly subkeplerian velocity. The equation of motion of a gas parcel in the radial direction is given by
 
\begin{equation}
r \Omega^2_g = \frac{GM r}{R^3} + \frac{1}{\rho_g} \frac{\partial P}{\partial r} \, ,
\end{equation}

\noindent where $\Omega_g$ is the rotation frequency of the gas, $M$ is the mass of the central object and $R$ is the distance of the gas parcel from this object. Assuming $P = c^2_g \rho_g$, this gives the well known relation for the gas orbital velocity $v_{\phi,g}$ \citep[see e.g.,][]{weid77}
 
\begin{equation}
v_{\phi,g} \equiv  v_K - \eta v_K \approx v_K + \frac{1}{2} \frac{c^2_g}{v_K} \frac{\partial \ln P}{\partial \ln r} \, ,
\end{equation}

\noindent where $v_K$ is the keplerian orbital velocity and $\eta$ is a measure of the gas pressure support.
 
Using the above relations, we derived the velocity of the gas in the radial direction from the azimuthal momentum equation of the viscous gas:

\begin{equation} \label{momentum}
\rho_g v_{r,g} \frac{\partial}{\partial r} ( r v_{\phi,g} ) = \frac{1}{r} \frac{\partial}{\partial r} ( r^2 T_{r\phi}) + \frac{\partial}{\partial z}(r T_{\phi z}) \, .
\end{equation}

\noindent where $T_{r\phi}$ and $T_{\phi z}$ are the shear stresses expressed as \citep[e.g.,][]{tak02} 

\begin{equation}
\begin{array}{ccc}
T_{r\phi} = r \nu \rho_g \frac{\partial \Omega_g}{\partial r} & \mathrm{and} & T_{\phi z} = r \nu \rho_g \frac{\partial \Omega_g}{\partial z} \, .\\
\end{array}
\end{equation}
 
\noindent Equation~\ref{momentum} directly yields the expression for the radial velocity of the gas :

\begin{equation}\label{velocity}
v_{r,g}(z) = \left[ \frac{\partial}{\partial r} (r^2 \Omega_g) \right]^{-1} \\
\left[ \frac{1}{r\rho_g} \frac{\partial}{\partial r} \left( r^3 \nu \rho_g \frac{\partial\Omega_g}{\partial r} \right) + \frac{r^2 \nu}{\rho_g} \frac{\partial}{\partial z} \left( \rho_g \frac{\partial\Omega_g}{\partial z}\right) \right]
\end{equation}

\noindent where we used the fact that $v_{\phi,g} = r \Omega_g$ and replace the shear stresses by their expressions.

Using the assumption of vertical hydrostatic equilibrium for the gas, its density is given by:

\begin{equation}
\rho_g(r,z) = \rho_0(r) e^{-\frac{z^2}{2H^2_g}} \, ,
\end{equation}

\noindent with

\begin{equation}
\rho_0(r) =  \frac{\Sigma_g}{\sqrt{2\pi} H_g} \, ,
\end{equation}

This set of equations allows us to determine the radial velocity of the gas flow as a function of the distance to the planet and height above the disk midplane. Note that the density-weighted average of equation~\ref{velocity} over $z$ results in the mean accretion flow velocity $v_\mathrm{acc}$ derived by \citet{lyn74} :
\begin{equation}
  v_\mathrm{acc} = - \frac{3}{\Sigma_g r^{1/2}}\frac{\partial}{\partial r} (\nu \Sigma_g r^{1/2}) \,.
\end{equation}

\begin{figure}
\includegraphics[scale = 0.45]{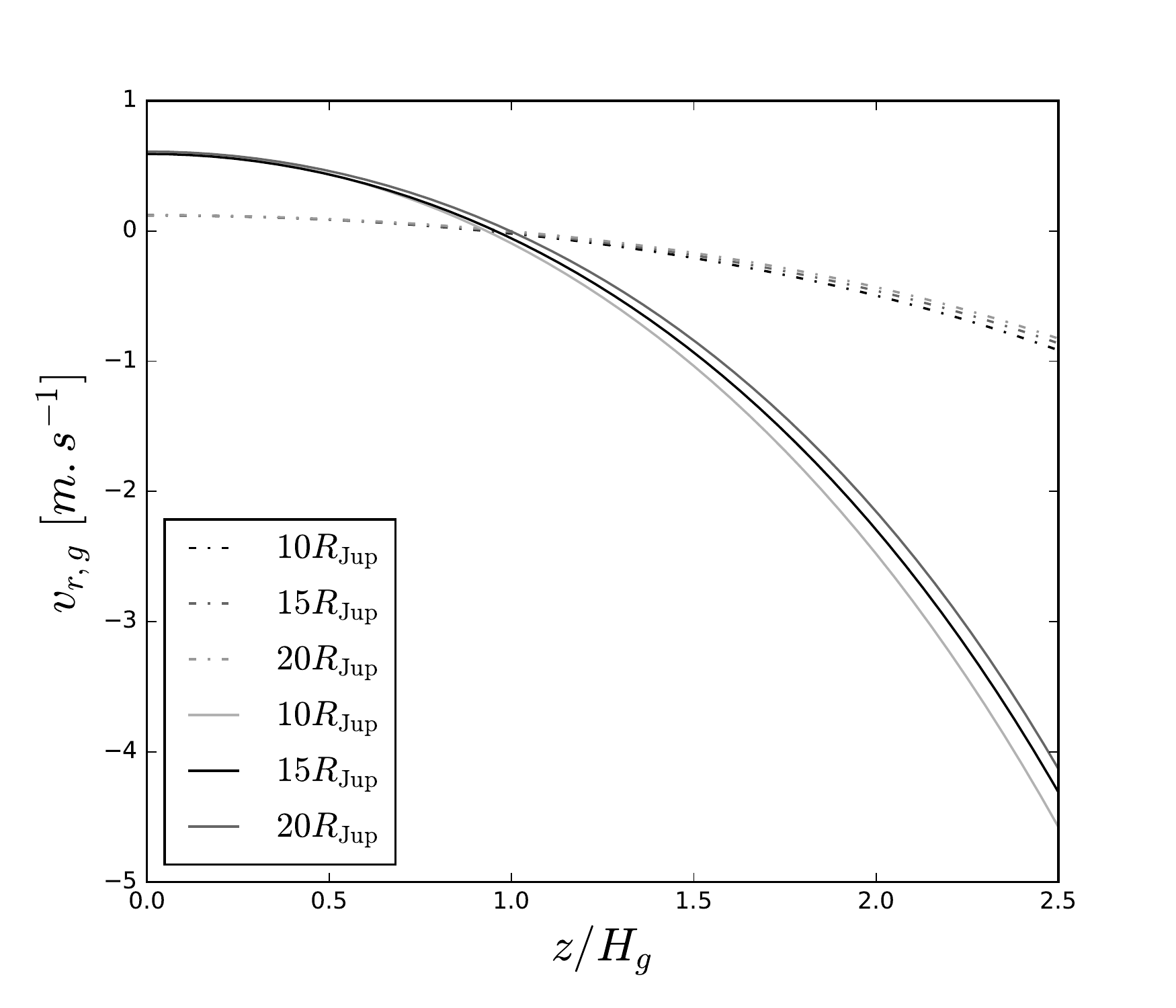}
\caption{Radial velocity profiles of the gas as a function of the height above the midplane at different distances from Jupiter. Solid and dashed lines correspond to profiles calculated with $\alpha=1\times 10^{-4}$ and $5\times 10^{-4}$, respectively.}
\label{gas_vel}
\end{figure}
 
Figure~\ref{gas_vel} represents the radial velocity vertical profiles calculated at different distances from Jupiter and for different values of $\alpha$. The velocity profiles are poorly influenced by the distance from the central planet. Instead, they strongly depend on the disk's viscosity where higher levels of turbulence result in larger velocities (both inward and outward) and consequently faster evolution of the disk. The velocities are small and slightly positive (outward) close to the midplane while at greater heights, namely in the less dense parts of the disk, they become larger and negative (inward). Such profiles have already been detailed in several studies of protoplanetary disks (PPDs) \citep[e.g.,][]{tak02,kel04,cie09}. It should be noted that such velocity profiles have not been found in turbulent simulations of disks \citep{fro11} because the Magneto-Rotational Instability (MRI), which is the source of turbulence in this simulations, results in non-uniform effective viscosity in the vertical direction. However, the outward radial velocity in the midplane of the CPD has been evidenced in several 3D hydrodynamic simulations \citep{tan12,kla06} as well as in MHD simulations \citep{gre13}. Moreover, only small dust grains that are well coupled with the gas can be significantly affected by its meridional circulation. The dynamics of larger grains/solids are mostly dictated by the deviation from keplerian orbital velocity of the gas (see Section~\ref{PDT}). It is therefore unclear, given the current knowledge of the structure of CPDs and PPDs, whether or not the radial velocity profiles we used are realistic, but this should hardly change our general conclusions.
  
 \subsection{Particles dynamics and thermodynamics}
 \label{PDT}
 
A lagrangian integration method is used to track the individual particles within the CPD. The transport model includes several mechanisms. Among them, the primary mechanism dictating the dynamical evolution of solids is the gas drag. Contrary to gas, solid particles are not pressure supported and their velocity do not deviate from the keplerian velocity. Solids consequently orbit around the planet faster than the gas does and feel a headwind. They transfer angular momentum to the gas via friction forces on a timescale called the stopping time of the particle $t_s$. This quantity generally depends on the size of the particle $R_s$, the gas density and the relative velocity $v_{rel}$ between the particle and the gas. Assuming that solids are spherical particles, their stopping time is \citep{per11,gui14}:
 
\begin{equation}
t_s = \left( \frac{\rho_g v_{th}}{\rho_s R_s} \mathrm{min}\left[ 1, \frac{3}{8} \frac{v_{rel}}{v_{th}} C_D(Re)\right] \right)^{-1}
\end{equation}
 
\noindent where $v_{th} = \sqrt{8/\pi}c_g$ is the gas thermal velocity, $\rho_s$ the density of the solid particle, assumed to be 1 g cm$^{-2}$ regardless of its size. The dimensionless drag coefficient $C_D$ is a function of the Reynolds number $Re$ of the flow around the particle \citep{per11}:
 
\begin{equation}
C_D=\frac{24}{Re} (1+0.27Re)^{0.43}+0.47\left(1-e^{-0.04Re^{0.38}}\right).
\end{equation} 
 
\noindent The Reynolds number is given by \citep{sup00}:

\begin{equation}
Re = \frac{4 R_s v_{rel}}{c_g l_g} .
\end{equation}

\noindent where $l_g$ is the mean free path of the gas.

The stopping time is divided into two regimes. The Epstein regime is valid when the particle size is smaller than the mean free path of the gas. In this case, the stopping time does not depend upon the relative velocity between the particle and the gas. When the particles are larger than the mean free path of the gas, the gas should be considered as a fluid. In such a case, the stopping time depends upon the relative velocity and the Reynolds number of the flow. In the limit $Re\ll 1$ \citep{gui14}, the conditions of the widely used Stokes regime are fulfilled. 
 
The equation of motion of the particles within the CPD is then given by:

\begin{equation} \label{eqmo}
\frac{d \bm{v}_s}{dt} = - \frac{GM_p}{R^3}\bm{R} - \frac{1}{t_s} (\bm{v}_s-\bm{v}_g).
\end{equation}

\noindent where $M_p$ is the mass of the central planet (here Jupiter), $\bm{R}$ the position vector of the particle, $\bm{v}_s$ its velocity vector and $\bm{v}_g$ is the velocity of the gas. The equation is integrated with an adaptive time step ODE solver\footnote{The ODE solver is available at the following webpage: \url{https://computation.llnl.gov/casc/odepack/}} \citep{bro89}, using Adams methods for particles with sizes down to $10^{-3} \, \mathrm{m}$. An implicit backward differentiation formula scheme is used to integrate the motion of lower size particles whose small stopping times imply a too restrictive time step for an explicit scheme (the time step should be smaller than the stopping times of the particles).

Small dust grains ($\sim$$\mu$m) have very short stopping times (e.g., $t_s \ll \Omega_K^{-1}$), meaning that they quickly become coupled with the gas. On the other hand, large planetesimals (tens or hundreds of kilometers in radius) have long stopping times ($t_s\gg \Omega_K^{-1}$) and their motion is hardly affected by the friction with the gas. Intermediate planetesimals, with sizes in the $\sim$cm range, efficiently loose angular momentum but on timescales that are too long to allow them to become coupled with the gas. These bodies thus always feel a headwind and they continue loosing angular momentum, causing them to rapidly drift inward towards the central planet. The solids that experience the most rapid inward drift are those whose Stokes number $S\!t$, namely the stopping time multiplied by the local keplerian frequency $(\Omega_K t_s)$, is of order unity.
 
\begin{figure*}
\centering
\includegraphics[width=\textwidth]{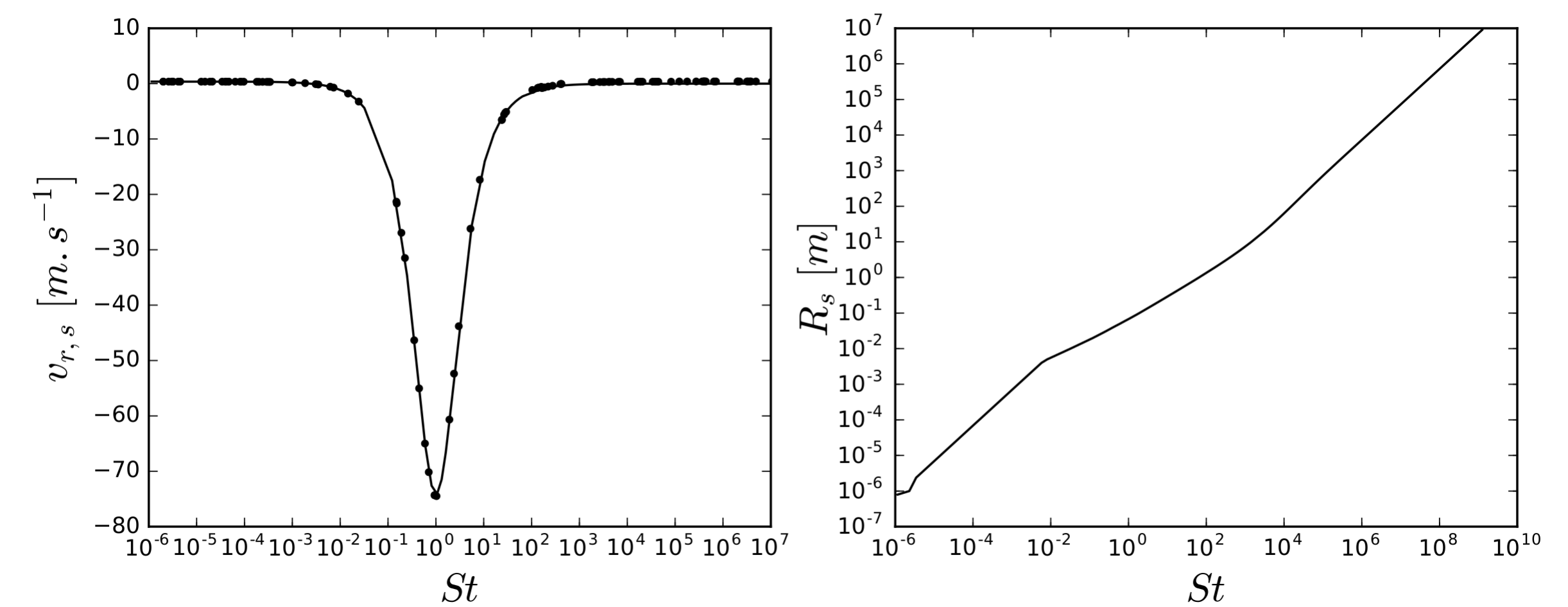}
\caption{\textit{Left}: particles radial velocity as a function of their Stokes number (black dots) at $15\,R_\mathrm{Jup}$ from a Jupiter mass planet in the midplane of a CPD with $\dot{M}_p = 10^{-7}\,M_\mathrm{Jup}\,yr^{-1}$ and $\alpha = 10^{-3}$. The solid line shows the solution of the analytical formula given by Equation~\ref{an} which fits well the results of our integration. Small dust grains with sizes smaller than $\sim10^{-3}\,\mathrm{m}$ have a slightly positive velocity which is that of the gas at the midplane ($v_{r,g}\simeq0.15\, \mathrm{m\,s}^{-1}$). Overall, there is more than one order of magnitude difference between the velocity of pebbles (solids with $St\sim1$) and those of the larger ($St\gg 1$) and smaller ($St \ll 1$) particles. \textit{Right:} correspondance between the Stokes number and the size of the particles.  }
\label{radvel}
\end{figure*}

Figure~\ref{radvel} represents the mid plane radial velocity of particles as a function of their Stokes number (left panel) as well as the size associated with the Stokes number (right panel) for solids at a distance of 15 $R_\mathrm{Jup}$ from Jupiter.The left panel of Figure~\ref{radvel} shows a comparison of the velocity of particles in the simulation (black dots) with that derived from the analytical formula \citep[see e.g.][]{birn12} :
\begin{equation}\label{an}
  v_{r,s}=-\frac{2St}{1+St^2}\eta v_K + \frac{1}{1+St^2}v_{r,g}.
\end{equation}
 Almost all solids are steady in the disk compared to the very rapid dynamics of the pebbles (particles with $S\!t \sim1$) that drift inward at high velocities.

The other mechanism affecting the motion of solids is turbulent diffusion. Turbulent eddies can entrain particles during their cohesion timescale and would efficiently mix radially and vertically small dust grains that couple well with the gas.
 The motion of solids due to turbulence is modeled following \citet{cie10,cie11} with a stochastic kick in the position of the particle \citep[see also][]{char11}. Additional advection terms are also added to account for the non uniform background gas density and diffusivity of solids (see eq.~\ref{adv}). For a detailed description of this kind of model we refer the reader to the work of \citet{cie10,cie11} and \citet{char11} who comprehensively describe the physcics modeled and demonstrate how the Monte Carlo method is able to solve for the advection-diffusion equation of the solids. Accounting for all transport mechanisms, the new position of a solid particle along any axis of a cartesian coordinate system after a timestep $dt$ can be expressed as \citep{cie10,cie11,char11}

\begin{equation}
x(t+dt) = x(t) + v_{adv} dt + R_1 \left[ \frac{2}{\sigma^2} D_p dt \right]^{\frac{1}{2}},
\end{equation}

\noindent where $x$ stands for any cartesian coordinate, $R_1 \in [-1;1]$ is a random number, $\sigma^2$ the variance of the random number distribution, $D_p$ the diffusivity of the solid particle and $v_{adv}$ is the term accounting for the non uniform density of the gas in which the particles diffuse as well as the non uniform diffusivity of the particles, and the forces experienced by the particle, namely the gravitational attraction from the central planet and the gas drag (see eq.~\ref{adv}). $D_p$ is related to the gas diffusivity through the Schmidt number $S\!c$ as \citep{you07}:

\begin{equation}
Sc \equiv  \frac{\nu}{D_p} \sim 1 + \frac{S\!t^2}{4},
\end{equation}
 
\noindent implying that solids with large Stokes number are not significantly affected by turbulence. The advective velocity $v_{adv}$ is given by \citep{cie10,cie11,char11}:

\begin{equation}\label{adv}
v_{adv} = \frac{D_p}{\rho_g}\frac{\partial \rho_g}{\partial x} + \frac{\partial D_p}{\partial x} + v_{s,x},
\end{equation}

\noindent where the two first terms account for the gradients in gas density and solid diffusivity and the last term is the velocity of the particle determined from its equation of motion (eq.~\ref{eqmo}).

We have also included the sublimation of water ice in our model to track the evolution of the ice fraction of the solids during their transport within the CPD. This ice fraction is compared with the present water content of the Galilean satellites. The surface temperature of the solids is calculated following the prescription of \citet{dan15}, in which several heating and cooling mechanisms are considered. The main heat source is the radiation from the ambiant gas at the local temperature $T_d$. Friction with the gas also heats up the surface of the body. Water ice sublimation on the other hand is an endothermic process that substantially lowers the temperature of the solid.
  
Finally, energy is radiated away from the surface at the surface temperature of the body. Taking into account all the heating and cooling sources, and considering that these processes only affect an isothermal upper layer of thickness $\delta_s$, the evolution of the surface temperature $T_s$ of the solid is given by \citep{dan15}

\begin{equation}
\begin{split}
\frac{4}{3} \pi \left[ R^3_s - ( R_s - \delta_s )^3 \right]  \rho_s C_s \frac{dT_s}{dt} = \frac{\pi}{8} C_D \rho_g R^2_s v^3_{rel} \\
+ 4\pi R^2_s     \epsilon_s \sigma_{SB} \left( T^4_d - T^4_s \right) \\
 + L_s \frac{dM_s}{dt}, 
\end{split}
\label{temp}
\end{equation}

\noindent where $R_s$ is the radius of the particle, $C_s$ is the specific heat of the material set to $1.6\times 10^3 \,\mathrm{J\,kg^{-1}\,K^{-1}}$ (specific heat of water ice at $\sim$200$\,$K), $\epsilon_s$ is the emissivity of the material, $\sigma_{SB}$ is the Stefan-Boltzmann constant and $L_s$ is the latent heat of sublimation of water ice $(L_s = 2.83\times 10^6 \, \mathrm{J\,kg^{-1}})$. Usually, the heating due to gas friction has a negligible effect so that the surface temperature of the bodies tends to equal that of the disk when water ice sublimation is not significant. On the other hand, when sublimation is important, the surface temperature can be significantly lowered (see Section~\ref{res} and Figure~\ref{surftemp} for more details).
    
The resulting mass loss rate due to water ice sublimation is then given by

\begin{equation}
\frac{dM_s}{dt} = -4\pi R^2_s P_v(T_s) \sqrt{ \frac{\mu_s}{2\pi R_g T_s}},
\label{abl} 
\end{equation}

\noindent where $P_v(T_s)$ is the equilibrium vapor pressure of water over water ice at the temperature $T_s$, $\mu_s$ the molecular weight of water and $R_g$ the ideal gas constant. The above expression is neglecting the effect of the partial pressure of water and holds in vacuum. In practice, $P_v$ should be replaced by $(P_v(T_s) - P_\mathrm{H_2O}(r))$ in Equation~\ref{abl}, with $P_\mathrm{H_2O}(r)$ the partial pressure of water vapor in the disk. However, we do not follow the evolution of the water vapor in this study and the initial composition of the CPD is uncertain as water was most likely in condensed form at Jupiter's orbit. Our expression therefore yields to "colder" snowlines as the sublimation of water ice should be inhibited whenever $P_\mathrm{H_2O} > P_v$ in more realistic conditions. The equilibrium vapor pressure $P_v(T_s)$ is computed from \citet{fray09} :
\begin{equation}
  \ln\left(\frac{P_v(T)}{P_t}\right) = \frac{3}{2}\ln\left(\frac{T}{T_t}\right) + \left(1-\frac{T_t}{T}\right)\gamma\left(\frac{T}{T_t}\right)
\end{equation}
\begin{equation}
  \gamma\left(\frac{T}{T_t}\right) = \sum\limits_{i=0}^6 e_i \left(\frac{T}{T_t}\right)^i
\end{equation}
\noindent where $P_t = 6.11657\times10^{-3}$ bar and $T_t = 273.16$ K are the pressure and temperature of the triple point of water respectively. The coefficients $e_i$ are given in Table~\ref{coeff}.

The thickness of the isothermal layer is given by \citet{dan15} as
\begin{equation}
\delta_s = min \left[ R_s , 0.3 \frac{K_s}{\sigma_{SB}T^3_s} \right]
\end{equation}
\noindent where $K_s$ is the thermal conductivity of ice ($\sim3\; \mathrm{W\,m^{-1}\,K^{-1}}$ at $200\,$K). At a surface temperature of 150 K the thickness of the isothermal layer is $\delta_s \sim 4.7\,$m while at 200 K it is reduced to $\sim$2 m. For the sake of simplicity, we do not consider here a mixture of ice and rock that would primarily have a slightly lower specific heat and a slightly higher thermal conductivity. The impact on our results focusing on the sublimation of water ice only would be minor as \citet{dan15} demonstrated that the differences in the ablation rates among completely icy and mixed composition bodies are no more than $\sim$10\%.
   
The equations depicting the surface temperature evolution and mass ablation rate are integrated together with the equation of motion of the particle. The change in radius caused by ice ablation is taken into account during the determination of the stopping time and consequently in the motion equation of the particle. For the sake of simplicity, we assume that the density of the solids is not modified during ice ablation and the radius of the particle is therefore always given by $Rs = (3M_s/4\pi\rho_s)^{1/3}$. This is equivalent to considering that the porosity of the body increases when ice sublimates. 

\section{Results}\label{res}

\begin{figure*}
\includegraphics[width=\linewidth]{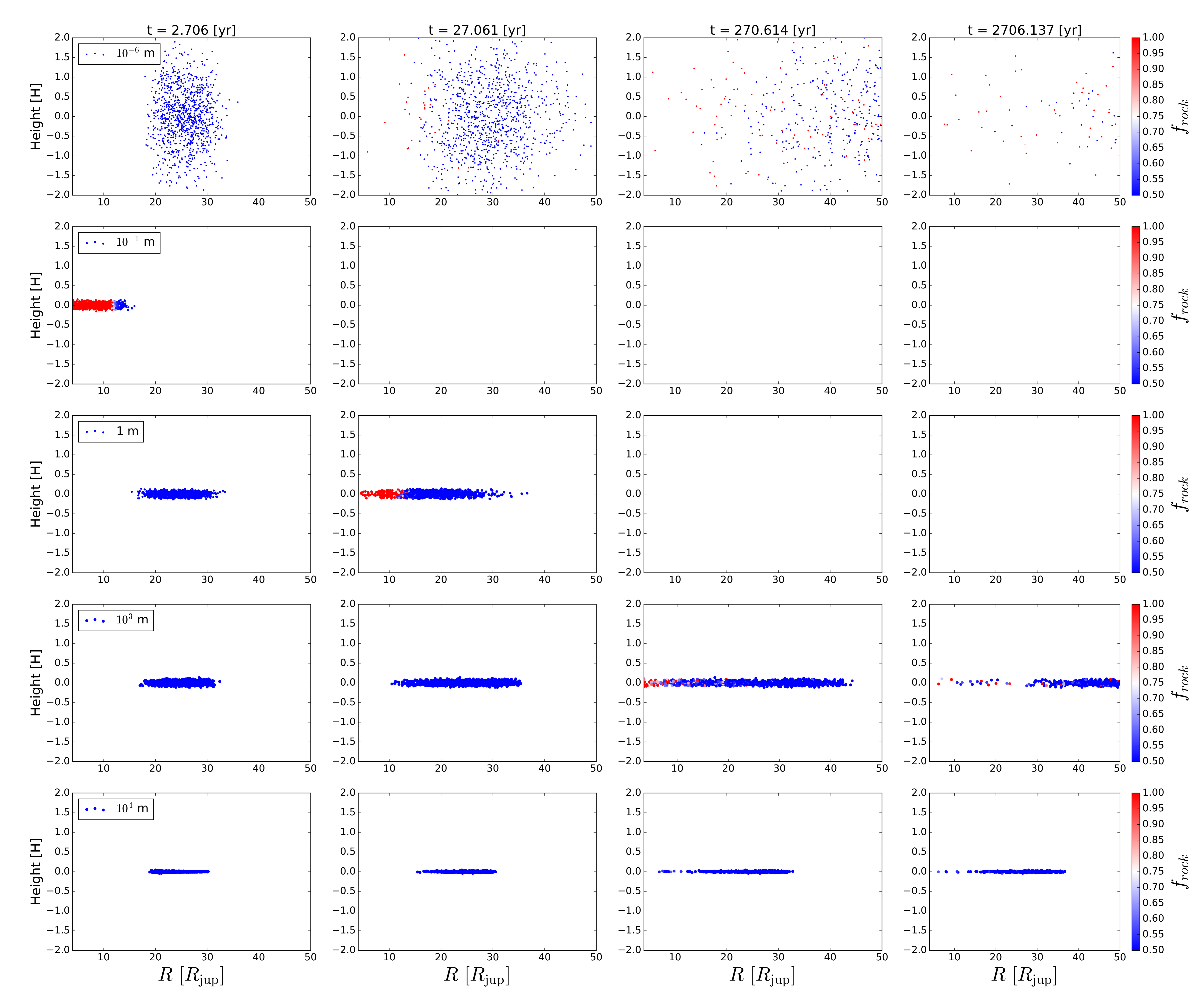}
\caption{From left to right: snapshots of the evolution of particles at different times within a Jovian CPD with parameters $\dot{M}_p = 10^{-7}\;M_\mathrm{Jup}\,yr^{-1}$ and $\alpha = 10^{-3}$. From top to bottom, each row displays the evolution of solids of different initial sizes with radii of $10^{-6}$, $10^{-1}$, 1, $10^3$ and $10^4$ m. The radial and vertical positions of the solids are expressed in $R_{\rm Jup}$ and local gas scale height respectively. The color of each particle illustrates its composition with bluer particles having a higher water ice mass fraction.}
\label{global}
\end{figure*}

Figure~\ref{global} presents the results of simulations with initial sizes of $10^{-6}$, $10^{-1}$, 1, $10^3$ and $10^4$ m, to illustrate their very different behavior in terms of dynamics and thermodynamics. 

We applied our model to particles of different initial sizes ($10^{-6}$, $10^{-1}$, 1, $10^3$ and $10^4$ m) and tracked the dynamical and compositional evolution over a short timespan (2700 years). Specifically, one thousand particles per size bin were initially released in the midplane of the CPD at distances ranging between 20 and $35 \, R_\mathrm{Jup}$. At the beginning of the simulation, all particles have an ice mass fraction $f_\mathrm{ice}=m_\mathrm{ice}/m_\mathrm{tot}=0.5$. The CPD is assumed to be in steady-state with $\dot{M}_p = 10^{-7} \; M_\mathrm{Jup}\,yr^{-1}$ and $\alpha = 10^{-3}$ which gives the surface density and temperature profiles drawn on Figure~\ref{subneb}, allowing to focus the results on solids' evolution. The inner edge of the disk is set equal to $3.5 \, R_\mathrm{Jup}$. Solids crossing this distance are considered lost to the planet, implying that their motion is no longer integrated. In Figure~\ref{global}, we display the rock mass fraction ($f_\mathrm{rock}=1-f_\mathrm{ice}$), height and distance to Jupiter of the solids as a function of time. 

The different dynamical behavior as a function of particle size is well illustrated in Figure~\ref{global}. A common feature for all particle sizes is the much faster vertical than radial diffusion timescale. The first column of the figure, showing the radial and vertical position of the solids after $\sim$2.7 years of evolution, illustrates the fact that solids are already distributed vertically and this distribution does not significantly change further in time. As expected, larger solids concentrate more in the midplane of the disk whereas micron sized dust particles are efficiently entrained by turbulence and follow the distribution of the gas. It is important to note that the vertical position of the solids (Figure~\ref{global}) is represented in units of the gas scale height $H_g(r)$ at the radial position of the particle. The radial drift of the particles also follows a well-known trend with very small particles (micron-sized) being well coupled with the gas, intermediate-sized particles (1cm-1m) drifting inward at a high pace, and large particles ($\geq$1 km) drifting inward and diffusing outward at a very low pace. 
 
Concerning the compositional evolution of the particles, some clear trends emerge (see Figure~\ref{global}). It appears that size strongly influences the ability of a given particle to retain water while drifting inward. In short, larger bodies are able to retain significantly more water than the smaller ones. For example, meter-sized bodies located inside of $\sim$$12 \,R_\mathrm{Jup}$ have lost all their water after 27 years of evolution whereas kilometer-sized bodies (fourth row of Figure~\ref{global}) have retained most of their water at the same location. The same applies for $10^3$ and $10^4$ m solids after 270 and 2700 years of evolution. It is also interesting to note that due to their limited inward drift and rather long sublimation timescales, water-free and water-rich kilometer-sized bodies can coexist at the same location, a feature that is not observed among the smaller particles.

The origin of such compositional evolution as a function of particle size is twofold. First, from Eq.~\ref{abl}, one can derive that the ablation timescale at a given location of a particle is $M_s (dM_s/dt)^{-1} \propto R_s$, implying that larger particles retain more water than smaller ones. Second, because water ice sublimation is an endothermic process, it cools down the surface temperature of large particles efficiently for longer time. Considering negligible the heating due to friction with the gas and that an equilibrium is rapidly attained, Equation~\ref{temp} reduces to
\begin{equation}
  \epsilon_s \sigma_{SB} (T_d^4-T_s^4) = L_s P_v(T_s)\sqrt{\frac{\mu_s}{2\pi R_g T_s}}.
  \label{eqtemp}
\end{equation}
When the release of sublimation heat is important (right-hand side of the equation), the surface temperature of the bodies departs from that of the surrounding gas.

\begin{figure}
\includegraphics[width=\linewidth]{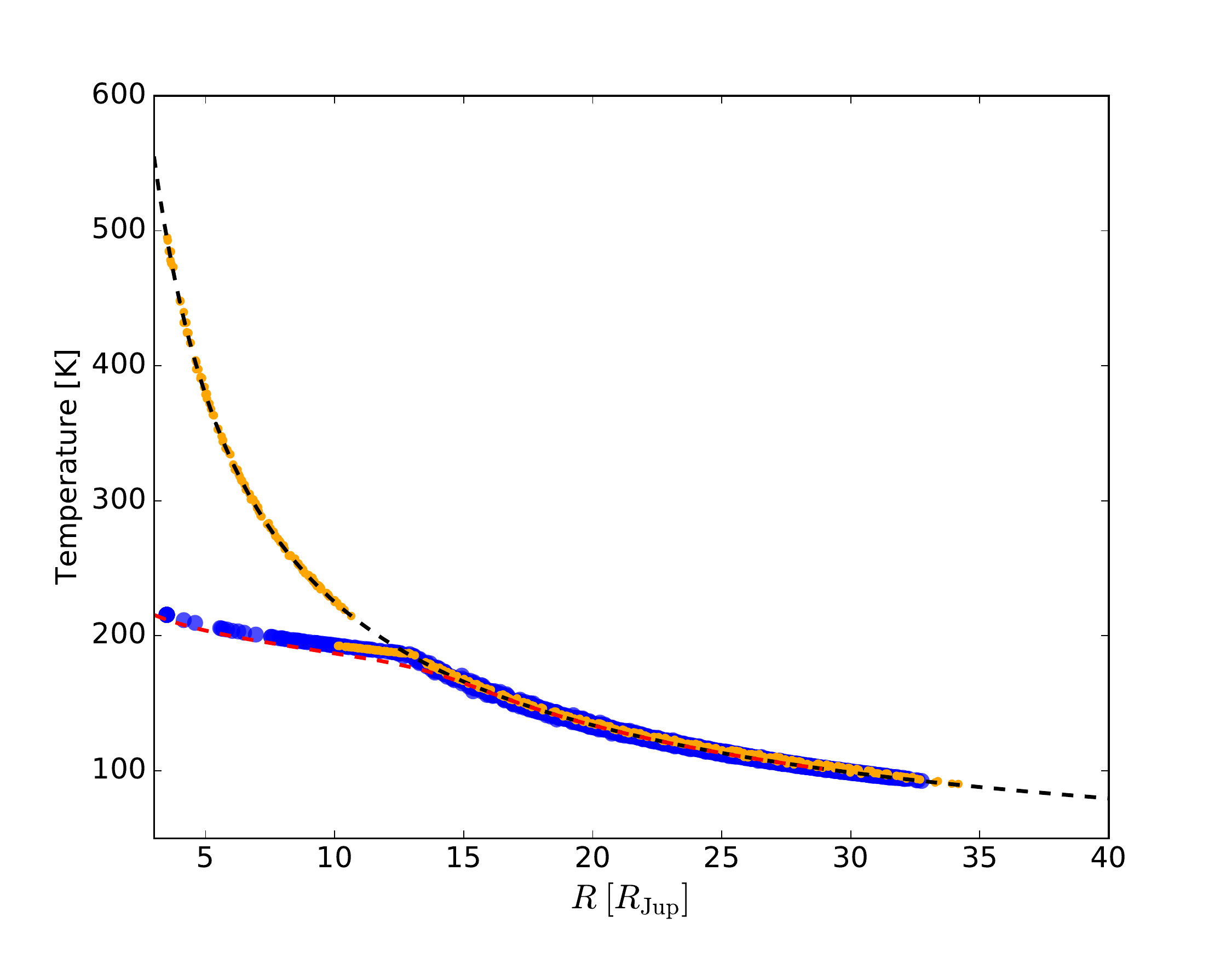}
\caption{ Surface temperature of 10 km (blue dots) and 10 cm (yellow dots) bodies as a function of the distance from Jupiter within a CPD with $\dot{M_p}=10^{-7}\,M_\mathrm{jup}\mathrm{yr}^{-1}$ and $\alpha=10^{-3}$. The black dashed line represents the temperature profile of the CPD while the red dashed line is the solution of Equation~\ref{eqtemp}. The high water ice ablation rates suffered by these bodies efficiently cools down their surface temperatures in the inner part of the disk, making them substantially depart from the ambient gas temperature. However, 10 cm bodies cannot retain water ice below $\sim$10$\,R_\mathrm{Jup}$ so that their surface temperature is that of the ambient gas interior to this distance.}
\label{surftemp}
\end{figure}

This process is well illustrated in Figure~\ref{surftemp} where the surface temperature of 10 km and 10 cm-sized planetesimals is shown (blue and yellow dots, respectively) along with the temperature of the surrounding gas (black dashed line) and the solution of Equation~\ref{eqtemp} (red dashed line). Closer to Jupiter, where the CPD is hotter, the temperature of these bodies departs from that of the gas because a significant amount of water sublimates at their surfaces. The surface temperature given by Equation~\ref{eqtemp} slightly underestimates the temperature but is a good approximation. In spite of that, the ablation timescale of 10 cm particles remains short and their water ice is entirely sublimated when they approach at distances $\lesssim 10\, R_\mathrm{Jup}$. Interior to this distance, the surface temperature of the 10 cm bodies abruptly catches up with the disk temperature. The efficient cooling during water ice sublimation and the fact that the sublimation timescale scales with the size of the object allows larger bodies to retain water over much longer timescales than their smaller siblings. 

 Due to the very short lifetime of the solids with sizes in the 10$^{-1}$-- 1 m range, we ran an other set of simulations to study in more details their evolution within the CPD. We also extended the size range down to 10$^{-2}$ m particles.

We ran simulations using 10,000 particles, released between 25 and 35 R$_\mathrm{Jup}$ and we opted to randomly re-inject in this region the particles that cross the inner edge of the CPD at 3.5 R$_\mathrm{Jup}$. In a way, we mimic a flux of pebbles that would originate from farther locations within the CPD. The parameters of the CPD are those used in the previous simulations. 

\begin{figure*}
\begin{center}
\includegraphics[width=\textwidth]{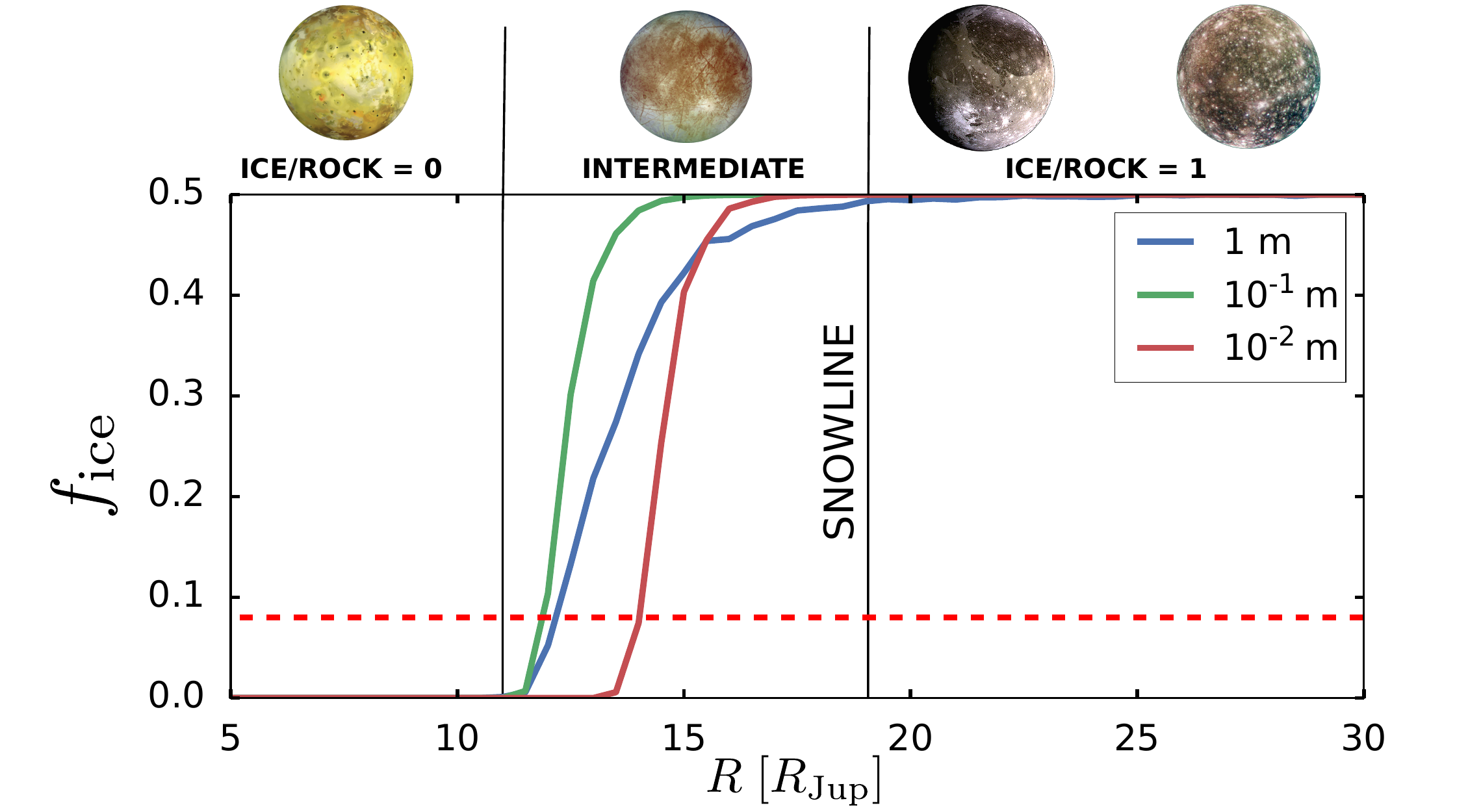}
\caption{Average water ice mass fraction of solids as a function of radial distance from Jupiter. $10^4$ particles of each size have been released in the 25--35 R$_\mathrm{Jup}$ region. The horizontal dashed line corresponds to Europa's estimated water mass fraction.}
\label{fice}
\end{center}
\end{figure*}
  
Figure~\ref{fice} shows the average water ice mass fraction $f_\mathrm{ice}$ of solids with sizes of $10^{-2}$, $10^{-1}$ and $1\,$m as a function of the distance to Jupiter. Due to the rapid dynamics of these solids and the fact that we re-inject them, an equilibrium is rapidly attained, meaning that the curves shown on Figure~\ref{fice} are steady in time for a stationary CPD. These curves would however shift towards Jupiter as the disk slowly cools down compared to the drift timescale of the pebbles. During their inward migration, solids gradually loose water ice and therefore exhibit a gradient in their water mass fraction as a function of the radial distance. The solids able to transport water the farthest inside the disk are the $10^{-1}$m pebbles because of their very rapid inward motion. The solids with a size of $10^{-2}$m display a very similar behavior although their water mass fraction is shifted in the outer radial direction. This shift is due to the shorter ablation timescale of $10^{-2}$m pebbles compared to that of the larger ones although their velocity is comparable (see Figure~\ref{radvel}). Larger meter sized bodies exhibit a less steep gradient in their water mass fraction because of their much slower inward velocities. They spend a greater amount of time in a given environment than smaller pebbles causing them to be more ablated and therefore they are not able to carry as much water as $10^{-1}\,$m pebbles.

Overall, we find that the pebbles define three distinct compositional regions. In the outer region, the solids mostly retain their primordial water content because they do not suffer from substantial sublimation. In the innermost region, the solids have already lost all of their water ice and are essentially rocky. In between these two regions, the particles exhibit a gradient in their water content over an area that is $\sim3 \; R_\mathrm{Jup}$ wide due to the combined effect of inward drift and sublimation.

\section{Discussion} \label{disc}

Here we put in perspective the results presented in the previous section with the current composition of the Galilean system. We try to provide some constraints on the size of the building blocks of the Jovian moons and discuss the implications on different mechanisms, such as the delivery of solids to the CPD or the migration of the satellites, which were not studied here. Overall, we try to provide new insights on the formation of the satellites of Jupiter and some exploration tracks for the future.

\subsection{Constraints on the size of the building blocks of the Galilean satellites}

We presented in Section~\ref{res} the dynamical and compositional evolution of particles with a wide range of sizes. We find that larger objects are able to retain more water ice than smaller ones, and that the ablation timescale of planetesimals with sizes $\gtrsim 10^4$ m is significantly enhanced in hot environments due to an efficient cooling of their surfaces.  
While it is common to assume that solids inside the snowline are rocky whereas the ones residing outward are icy \citep[e.g.,][]{alib05, sas10}, our results show that the solids embedded within Jupiter's CPD should have been (at least initially) relatively smaller than $10^3$ m to ensure this.
If the initial building blocks of the satellites were large (D$\geq$$10^3$ m) icy objects ($f_\mathrm{ice}$=1), Io and Europa would probably have formed with substantially more water than they possess today.
This finding is also supported by the study of \citet{dwy13} which demonstrated that water loss during collisions of large planetesimals is not a sufficient mechanism to account for the formation of a water free Io and Europa with less than 10\% water by mass. Conversely, if the initial building blocks of the satellites were small (D$\leq$$10^{-6}$ m) icy particles (ice/rock=1), Io and Europa would have formed without water and Europa should be dry today.  

There is only one size range that allows the direct formation of a dry Io, of a Europa with a low water content and of two icy moons (Ganymede, Callisto) in the outer region of the CPD, namely $10^{-2}$ m $\leqslant $ D $\leqslant 1$ m. If our proposed scenario is the right one, this implies that Europa could have had any water content between 0 and 50\% while forming in the intermediate region (see Figure~\ref{fice}). In summary, the growth of Europa could have been restricted to this ``intermediate" region, where the protosatellite would have accreted partially dehydrated, drifting material. Recent studies have shown that the accretion of solids with a Stokes number close to unity, such as those solids we present on Figure~\ref{fice}, is very efficient \citep[e.g.,][]{lam12}. These pebbles are therefore good building blocks candidates as their composition within the Jovian CPD could have defined three distinct compositional regions coherent with the current water content of the Galilean satellites.

It should be noted that the positions of the different regions defined on Figure~\ref{fice} do not match the current location of the Galilean satellites. Whereas it would be easy to adjust the mass accretion rate $\dot{M}_p$ to shift the position of the different regions, we do not want to suggest that these bodies formed in a steady disk or that they necessarily formed at the position we observe them today by doing so. These issues are further discussed in the next section.

\subsection{Caveats of the model and roadmap for future research}

We discuss here some of the processes that likely played a role during the formation of the satellites and that we did not study here, and how they would fit with our findings. We also recall the assumptions of the model we used and how it affects our results. 

\paragraph{Model assumptions}
   
We start here by discussing the assumptions upon which our results rely and some of the processes we neglected in this study. In our simulations, we considered that solids lose water via sublimation of water ice and that the refractory part remains. This gradual sublimation of pebbles gives rise to the region suitable for the formation of Europa like bodies. Other studies of grain sublimation suggest that solids are disrupted into small micrometer dust grains when they cross the snowline \citep[see e.g.,][]{sai11}. In such a case, no gradient in the composition of the solids would exist, but rather a twofold population constituted of very small silicate grains inside the snowline and large icy grains outside. Whether or not disruption of the grains occurs at the snowline depends on the structure of the grains. Very porous aggregates of silicate monomers covered with ice are prone to disruption while more compact aggregates or collisional fragments of larger bodies would more likely stay intact. The structure of the solids embedded within the Jovian CPD is uncertain and would primarily depend on the delivery mechanism of solids within the CPD, which is discussed next.

 In addition, our model does not consider the condensation of water vapor onto grains. Although the effect on centimeter or larger grains should be moderate, it has a great importance for the evolution of small dust grains onto which condensation will preferentially occur \citep{ros13}. As we did not include grain growth either in our model, we miss effects such as local water vapor or solids enhancement close to the snowline \citep[e.g.,][]{cie06}. This is however beyond the scope of this study and would only be relevant if the solids within the CPD built-up from small dust grains. This depends, again, on wether solids are primarily brought to the CPD in the form of small, well coupled grains or in the form of larger, already decoupled aggregates. As we discuss below, this question remains to be investigated but our results would be more consistent with the delivery of already decoupled and rather compact solids.
 
\paragraph{Delivery mechanism of solids}
  
The main origin of solids in the Jovian CPD, which is deeply connected with the formation of Jupiter, is still debated. Two different mechanisms have been proposed to feed the CPD. \citet{can02} proposed that small dust grains that couple with the gas are entrained with the inflow onto the CPD whereas \citet{mos03a,mos03b} and \citet{est06} argued that larger planetesimals crossing Jupiter's orbit could be captured through gas drag within the CPD.

While the first mechanism has not been quantitatively studied, we can note some important caveats. It is expected that dust grains can grow up to decoupling sizes with Stokes numbers $\gtrsim 10^{-2}$ in the regions where the giant planets formed and that the population of larger grains carry most of the mass \citep[e.g.,][]{birn11,birn12}. This is also required to rapidly grow the cores of giant planets through pebble accretion \citep{lam14a}. It results that most of the solids mass should reside close to the midplane of the PPD in decoupled solids. This is hard to reconcile with the view of \citet{can02} who advocated fiducial dust-to-gas ratio of 1\% in the Jovian CPD. It is more likely that the gas accreted by Jupiter and its disk, which proceeded through the heights of the PPD as demonstrated by 3D hydrodynamic simulations \citep{tan12,szu16}, was depleted in dust. Interestingly, this depletion in dust benefits to giant planets formation as this would substantially reduce the opacity of their envelope, allowing a much faster contraction of the envelope and triggering runaway gas accretion more rapidly \citep{lam14b,bits15}.

Concerning the second mechanism, \citet{est06} and \citet{mos10} argued that at the time of the formation of the satellites, planetesimals in heliocentric orbits would have their eccentricities and inclinations excited by almost completely formed nearby giant planets. Collisions among these excited planetesimals would have led to intense collisional grinding and resulting bodies in the meter to kilometer size range \citep{char03}. This would provide suitable conditions for the capture of planetesimals by the CPD as their high inclinations and eccentricities would  place them onto Jupiter crossing orbits. The capture of this collisional fragments in the meter to kilometer size range is more in line with our study than the inflow of small grains, and as we mentioned, with the timing required to accrete Jupiter's envelope. It is also in agreement with the fact that starting out with large icy bodies (tens or hundreds of kilometers) would lead to the formation of hydrated inner satellites as neither collisions, as demonstrated by \citet{dwy13}, nor sublimation, as we pointed out in this study, seem efficient enough mechanisms to dehydrate such large building blocks. This capture scenario, however, remains to be investigated and quantified.
A better knowledge of the initial solids size and mass distribution within the Jovian CPD is crucial to disentangle from different formation mechanisms of the Galilean satellites. The key question here being to determine wether enough mass in the meter to tens of meter range can be brought within the CPD for pebble accretion to be relevant. 
 
\paragraph{Time evolution of the CPD and migration of the satellites}
  
Here we briefly discuss the time evolution and cooling of the CPD which we have neglected to focus on the evolution of the solids only. Although its structure and evolution timescale are very poorly constrained, the disk surrounding Jupiter likely evolved with time. Depending on the viscosity and mass accretion rate, the evolution of the CPD could have occurred on timescales ranging from $\sim10^4$ to $10^6$ years \citep{mig16}. 
The evolution and cooling of the CPD was therefore very slow compared to the inward drift of pebbles. It results that the composition of these solids would not be directly affected by the cooling of the disk. 
Provided that icy pebbles come from the outer parts of the CPD and drift towards its inner regions, they should always exhibit a gradient in composition when crossing the snowline. 
The disk's evolution would only affect the location of the snowline, namely the region where the gradient exists.

The question that remains to be elucidated is then wether or not the complete formation of the satellites, and particularly Europa, could have occurred in a given region matching their composition. 
This would depend on the ratio of their growth/migration timescale to the CPD's evolution timescale. 
Fully forming Europa in the region inside the snowline would either imply that i) its growth timescale was much faster than its migration timescale and the cooling timescale of the CPD or ii) its migration timescale was comparable to the disk evolution timescale so that Europa migrated inward together with the snowline as the CPD cooled over time.
While i) could be hard to reconcile with the fact that Callisto migth not be fully differentiated, implying a formation timescale of $\sim$10$^5$ years \citep[see e.g.,][]{can02}, several recent studies have shown that planet traps, i.e. regions where migration is halted, are associated with the water snowline inside PPDs \citep{bits15,bits16,bail15,bail16}, making scenario ii) a promising one. As the snowline moves inward over time, so does the migration trap, offering the possibility to tie the migration of a body to the evolution of the snowline.

\section{Conclusions} 
\label{end}

In this study, we have shown that the overall bulk composition of the Galilean satellites could be naturally accounted for in a pebble accretion scenario. The strong inward drift of these solids leads to the rapid emergence of well defined regions in terms of composition that can reproduce the gradient in water mass fraction existing among the satellites. The strongest implications of this scenario are the existence of pebbles that do not completely fragment when crossing the snowline and the fact that each satellite fully accreted in a given region. The latter imply that the migration of the satellites must have been somehow tied to the evolution of the snowline as its position determines the location of the different regions. Though it needs to be investigated in the case of the Jovian CPD, the existence of a relationship between migration and snowlines seems to be supported by recent theoritical developments about type I migration \citep[e.g.,][see also Section~\ref{disc}]{paar10,bits14}.
  
It is very delicate to determine whether the Jovian moons density gradient results directly from a gradient in the water mass fraction of the solids they accreted or from a more complicated interplay among their growth, migration and CPD's evolution given our current knowledge of these processes. While recent developments in 3D hydrodynamics simulations help better understand the accretion of gas onto the CPD \citep[e.g.,][]{tan12}, more realistic equations of state are needed to constrain the density and temperature of the CPD \citep[see e.g.,][]{szu16}. A better understanding of the formation and structure of circumplanetary disks and the delivery mechanisms of solids is crucial for further developments of Galilean satellites formation models. 

\acknowledgements
T.R. and O.M. acknowledge support from the A*MIDEX project (n\textsuperscript{o} ANR-11-IDEX-0001-02) funded by the ``Investissements d'Avenir'' French Government program, managed by the French National Research Agency (ANR). O.M. also acknowledges support from CNES.


\begin{deluxetable}{cc}
\tabletypesize{\scriptsize}
\tablecaption{Coefficients for the polynomial relation giving the equilibrium vapor pressure of water at a given temperature.}
\tablewidth{0pt}
\tablehead{
\colhead{$i$} & \colhead{$e_i$}}
\startdata
 0 & 20.9969665107897 \\
 {} & {} \\
 1 & 3.72437478271362 \\
 {} & {} \\
 2 & $-13.9205483215524$ \\
 {} & {} \\
 3 & 29.6988765013566\\
 {} & {} \\
 4 & $-40.1972392635944$ \\
 {} & {} \\
 5 & 29.7880481050215 \\
 {} & {} \\
 6 & $-9.13050963547721$ \\
\enddata
\tablecomments{The coefficients are taken from \citet{fray09}}
\label{coeff}
\end{deluxetable}

\end{document}